\documentclass[showpacs,floatfix,aps,prl,preprint,superscriptaddress]{revtex4-1}
\usepackage{graphicx}
\usepackage{xfrac}
\usepackage{amssymb}
\usepackage{tikz}
\usepackage{amsfonts} %maths
\usepackage{amsmath} %maths
\usepackage[utf8]{inputenc} %useful to type directly diacritic characters
\usepackage{mathtools}
\usepackage{xcolor,colortbl}
\usepackage{bm} %revtex4.1 for bold symbols
\usepackage{array}
\usepackage{color}  % STEFANO
\usepackage{relsize} % for AFLOWpy
 \usepackage[english]{babel}
 \usepackage{amssymb,amsmath}
 \usepackage{color}
 \usepackage{pstricks}
 \usepackage{bbold}
 \usepackage{graphicx}
 \usepackage{color}
 \usepackage{verbatim}
 \usepackage{fancyvrb}
 \usepackage{alltt}
 \usepackage{longtable}
 \usepackage{amsmath}

\definecolor{Blue}{rgb}{0.00,0.00,1.00}
\definecolor{Red}{rgb}{1.00,0.00,0.00}
\definecolor{Green}{rgb}{0.00,0.50,0.00}

\begin{document}

\title{Hyperbolic metamaterials with extreme mechanical hardness}

\author{Arrigo Calzolari,*}
\email {Email: arrigo.calzolari@nano.cnr.it}
\affiliation{CNR-NANO Research Center S3, Via Campi 213/a, 41125 Modena, Italy}
\author{Alessandra Catellani}
\affiliation{CNR-NANO Research Center S3, Via Campi 213/a, 41125 Modena, Italy}
\author{Marco Buongiorno Nardelli,}
\affiliation{Department of Physics, University of North Texas, Denton, TX 76203, USA\\
Center for Autonomous Materials Design, Duke University, Durham, NC 27708, USA}
\author{Marco Fornari}
\affiliation{Department of Physics and Science of Advanced Materials Program, Central Michigan University, Mt. Pleasant, MI 48859 USA\\
Center for Autonomous Materials Design, Duke University, Durham, NC 27708, USA}

% Keywords: Please provide a minimum of three and a maximum of seven keywords, separated by commas

\date{\today}

% Abstract should be written in the present tense and impersonal style (i.e., avoid we), and be at most 200 words long
\begin{abstract}
Hyperbolic metamaterials (HMMs) are highly anisotropic optical materials that behave as metals or as dielectrics depending on the direction of propagation of light. They are becoming essential for a plethora of applications, ranging from aerospace to automotive, from wireless to medical and IoT.
These applications often work in harsh environments or may sustain remarkable external stresses.  This calls for materials that show enhanced optical properties as well as tailorable mechanical properties.
Depending on their specific use, both hard and ultrasoft materials could be required, although the combination with optical hyperbolic response is rarely addressed. Here, we demonstrate the possibility to combine optical hyperbolicity and tunable mechanical properties in the same (meta)material, focusing on the case of extreme mechanical hardness.
Using high-throughput calculations from first principles and effective medium theory, we explored a large class of layered materials with hyperbolic  optical activity in the near-IR and visible range, and we identified a reduced number of ultrasoft and hard HMMs among more than 1800 combinations of transition metal rocksalt crystals. Once validated by the experiments, this new class of metamaterials may foster previously unexplored optical/mechanical applications.
\end{abstract}

\maketitle

\section{Introduction}
The ability to engineer new materials by integrating different components at the nanoscale has revolutionized science and technology in the last 100 years.
The transistor \cite{nobel56}, the giant magneto-resistance \cite{nobel07}, and the pioneering work of
Veselago and Pendry \cite{Veselago:67,0034-4885-68-2-R06} in the area of negative refraction have shown unexpected physics and opened the doors to new technologies \cite{Pendry00,Engheta02}.
Hyperbolic optical metamaterials (HMMs) fall within this scientific endeavor. They are  highly anisotropic optical materials which behave as metals  or as dielectrics depending on the direction of propagation of the electromagnetic field. Their properties are reflected in the sign of the diagonal terms of the dielectric tensor that is controlled during deposition by engineering  the composition of the multilayer system in the subwavelength regime \cite{Ferrari:2015ep}.
Due to their hyperboloidal isofrequency surface, HMMs have been proposed for a broad range of applications working from THz to UV-visible frequencies, including negative refraction \cite{Pendry06,Anonymous:2019er}, optical cavities \cite{Yango:20120ht}, biosensing \cite{Kabashin:2009kk}, and waveguides \cite{PhysRevB.73.155108}.
The hyperbolic behavior of materials is particularly relevant in the visible range for application such as superlenses \cite{Rho:2010ht}, cloaking \cite{Shalaev07},  sub-wavelength imaging \cite{Jacob:06, Alu:2006ud}, and perfect-absorbers for IoT \cite{Amiri:2020er}.
In addition, the potential for highly directional propagation of electromagnetic modes localized at subwavelength scales, e.g. volume plasmon polaritons \cite{PhysRevB.75.241402,Zhukovsky:13}, opens a new route into several nanophotonic applications as fluorescence engineering \cite{Krishnamoorthy12},
super-Planckian thermal emission \cite{Guo12}, subsurface sensing \cite{Taubner06}, cryptography \cite{Liu:2016fk}, and Dyakonov plasmons \cite{Narimanov08}.

The outstanding optical value of HMMs for applications cannot be disentangled by their mechanical properties, especially when the devices are expected to perform at extreme conditions such as high temperature, high pressure, under large tensile strain, and so on. For example, highly efficient antennas and radars are necessary for improved wireless communication, space vehicle navigation, and GPS satellites.
Antenna technology has gained great advantages from the integration with HMMs materials \cite{Monticone:2017cr} that enhance the dipole emission to free space, and provide a broadband impedance matching \cite{abbyfine:2020vh,Valagiannopoulos:2014ku}. However, in space conditions these devices need to withstand uniquely harsh environments: strength, hardness and stiffness become unavoidable properties to  resist to the forces and the collisions these systems are exposed to. Similar arguments hold for HMM-based devices for microwave \cite{Schuri2006dp} or optical \cite{Shalaev07}  cloaking exploitable in naval or military applications as well as for radar scanning systems for automotive \cite{Liu:2015dl}.
For all these applications the possibility to use {\em hard} (or {\em ultrahard}) materials could improve the mechanical resistance of the optical device.
On the opposite side, {\em ultrasoft} metamaterials could be of great interest for applications, e.g., in the medical industry  \cite{Sreekanth:2016gk}, where the ability to manipulate electromagnetic waves and provide diagnostic images can be coupled to the extremely flexibility of soft matter and the ability to  encapsulate and transport molecular systems (e.g.,  drugs) in biological systems.
On a general ground, the combination of hyperbolic optical properties and tailorable mechanical properties would allow to optimize the characteristics (e.g. endurance)  of the optical devices, opening to previously unexplored solutions.
In order to achieve this goal, the concept of optical metamaterials has to be extended to the mechanical realm \cite{Gao:2010do} by considering the effect of the multilayer geometry on the elastic properties and mechanical hardness.

Unfortunately, both {\em ultrasoft} and {\em hard} materials are rare and the ones known so far such as diamond, cubic BN, metal borides and carbides
(WB$_4$, ReB$_2$, WC, TiC), and ceramics (Si$_3$N$_4$, Al$_2$O$_3$), do not exhibit the special optical properties of HMMs  \cite{Yeung:2016hb}. This calls for new classes of
artificial materials.

By synergistically optimizing the optical properties and the mechanical responses, we have designed HMMs with extreme  ({\em ultrasoft} and {\em hard}) mechanical hardness starting from simple transition-metal rocksalt crystals, such as
TiN, ZrN, which are practical for the realization of hyperbolic superstructures \cite{Naik:2014hl}.
Our results are based on high-throughput first principles and  effective medium theory  calculations of the optical and mechanical  properties of periodic superlattices and provide an efficient strategy to  fine engineering  the optical and mechanical properties. We demonstrate the effectiveness of the approach by identifying several ultrasoft and hard HMMs among more than 1800 rocksalt combinations.

\begin{figure}
\begin{center}
\includegraphics[width=0.6\textwidth]{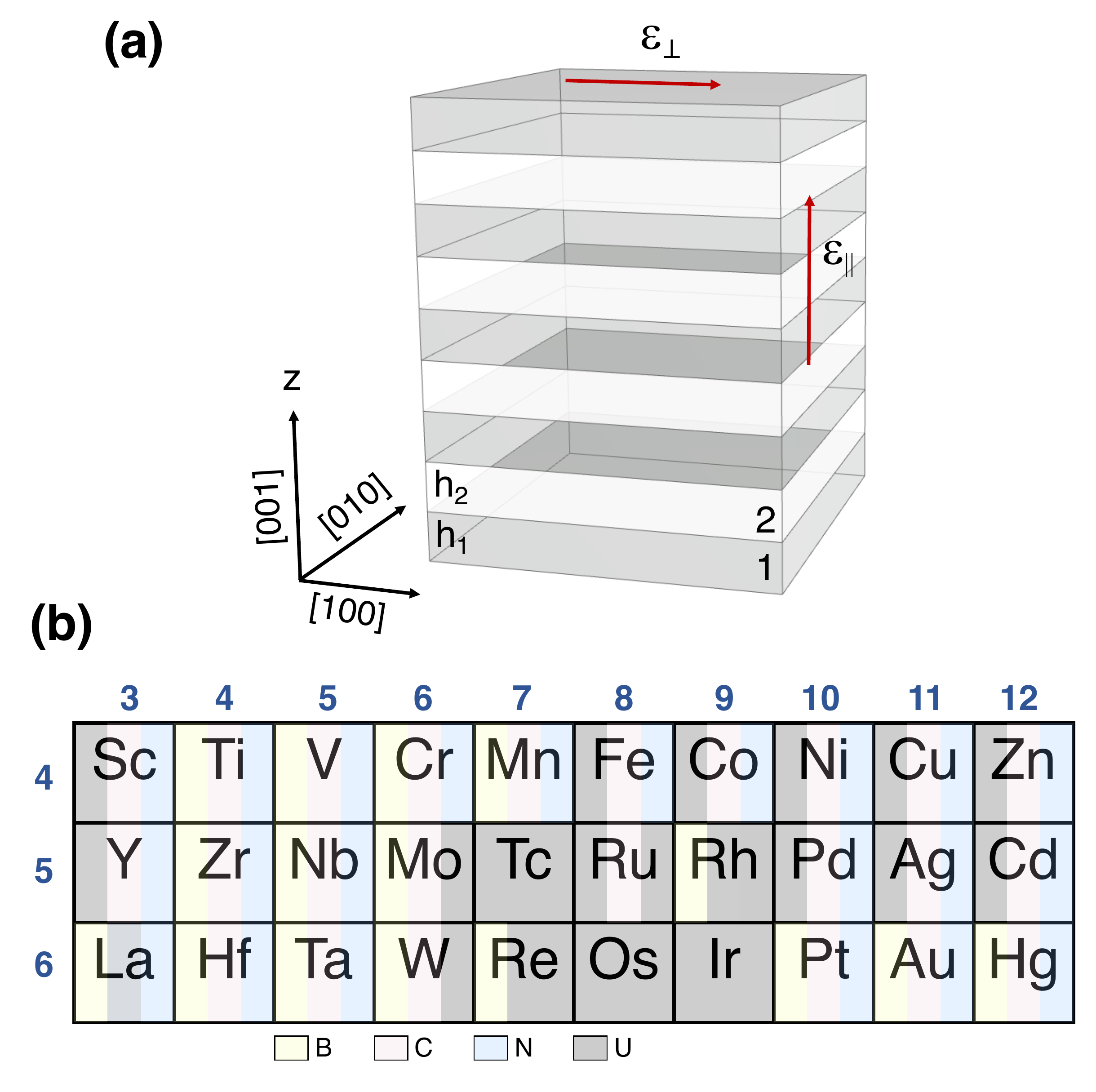}
\caption{a) Scheme of the superlattice models used to design the metamaterials.
Red arrows indicate the components of the dielectric function along the direction parallel ($\epsilon_{\parallel}$)
and perpendicular ($\epsilon_{\perp}$) to the optical axis $z$. High symmetry directions of cubic lattice are
aligned to cartesian axes. b) Table of TMX rocksalt compounds obtained combining the 30 transition metals (TM) with B, C, and N (X).
Gray sections indicate that the material is mechanically unstable (U).}
\label{fig1}
\end{center}
\end{figure}

%%%%%%%%%%%%%%%%%%%%%%%%%%%%%%
\section{Descriptors and selection criteria}
%%%%%%%%%%%%%%%%%%%%%%%%%%%%%%

{\em Single TMX crystals.} We considered superlattices composed of the alternating layers of two rocksalt TMX materials along the cubic [001] direction (Figure \ref{fig1}a). Ninety TMX compounds were obtained using  all possible transition metals (TM)
with the non-metal elements X=(B, C, N).  All structural, electronic, optical and elastic properties of single TMXs were evaluated from first principles, by using approaches based on density functional theory (DFT). The optimized  lattice parameter ($a_0$), the formation energy ($\Delta U$) with respect to the stable phases of the single components, and the complex dielectric function ($\hat{\epsilon}=\epsilon_r+i\epsilon_i$) were calculated with the Quantum Espresso code \cite{Giannozzi:2009p1507}. The ElaStic code  \cite{Golesorkhtabar:2013jy} has been used to obtain the elastic stiffness coefficients ($c_{ij}^{(\ell)}$) of the $\ell$-th TMX crystal; see Sec. 5 and Sec. S1 of Supporting Information (SI) for the full description of the method.

About one third of the systems (28/90) resulted to be thermodynamically and mechanically unstable
having both formation energies $\Delta U>0$ and the shear coefficient $c_{44}<0$.
MoN has negative formation energy but it has been discarded for its mechanical instability ($c_{44}<0$) in the cubic phase, in agreement with previous theoretical calculations \cite{PhysRevB.71.214103}.
The remaining 62 stable compounds formed the building blocks for the
superlattices. A summary of the rocksalt compounds with negative formation energy can be found in Figure \ref{fig1}(b).

Carbide and nitride compounds of the groups {4} and {5}  (e.g. TiC, TiN, VN, etc.) are stable and very well
characterized from both experimental \cite{2P-LPbIU,aIVZFYhh} and theoretical  \cite{Iuga:2007ck,2012JAP,Quesne:2018ca,PhysRevB.31.752} perspectives.
These are hard refractory materials exploited in a large range of applications from mechanics to electronics and optics. The carbide and nitride compounds with the remaining transition metals are more rare (e.g. RuC, AuC) and partially unexplored (e.g. FeC, MnN).
Monoboride compounds are also quite rare, being the diboride structures the most common ones. Most monoborides favor the orthorhombic or tetragonal phases \cite{Cardarelli:2008to}, although the metastable NaCl structures have been predicted \cite{Mohn:2000gh}.

All bulk components are metallic, except ScN, YN, and LaN that exhibit small indirect bandgaps ($E_g<1.0 eV$). The analysis of  the main electronic and optical properties of the stable TMX crystals is reported in Sec. S2 of SI. The stiffness coefficients, the elastic moduli, and the mechanical parameters  of all 62 crystals are summarized in Sec. S3 of SI, in very
good agreement with previous results (see SI for references).

{\em TMX superlattices.} By combining all possible pairs of stable TMX rocksalts, we constructed 1891 binary superlattices composed of alternating layers of different materials along the [001] direction (Figure \ref{fig1}a). The fractional volume of the $\ell$th layer is defined as $f_{\ell}=h_{\ell}/h$ ($\ell=1,2$), where $h=h_1+h_2$ is the volume of the superlattice unit.
The optical and elastic properties of the  superlattices were obtained within the effective medium theory (EMT) \cite{emt}.
Experimentally, the thickness of the individual layers ranges from 5 to 200 nm.
It is worth noticing, that although the effective medium theory used in this work  depends only on the volume ratio and not on the physical thickness of the single layers,
confinement and interface effects in ultrathin layers could have important consequences. However, previous experimental  \cite{Naik:2014hl} and theoretical results \cite{OMEX} demonstrated that EMT holds in the case of metallic layers (e.g. TiN) down to the shortest 5 nm thickness.
We have tested the validity of the EMT for selected configurations, as discussed in Sec. S4 of SI.

HMMs are anisotropic uniaxial materials that derive their name from the shape of the isofrequency surface $\omega({\bf k})$ where {\bf k}, and $\omega$ are the wavevector and the frequency of the radiation.
In non-magnetic uniaxial materials, the dielectric tensor can be written
in terms of two diagonal parameters: $\epsilon_{\parallel}$ and $\epsilon_{\perp}$,
where the subscripts $\parallel$ and $\perp$ indicate components parallel and perpendicular to the anisotropy axis (Figure \ref{fig1}).
If one of the two parameters is negative, the isofrequency surface opens up into a hyperboloid.
The choice $\epsilon_{\perp}>0$ and $\epsilon_{\parallel}<0$   corresponds to a two-sheet hyperboloid  and the medium is a
{type-I} metamaterial; the choice $\epsilon_{\perp}<0$ and $\epsilon_{\parallel}>0$  describes a one-sheet hyperboloid,
and the medium is called {type-II} metamaterial  \cite{Ferrari:2015ep,Poddubny:2013cy}.
Provided that the incident wavelength is large compared to the thickness of the constituent layers, the
dielectric tensor of the superlattice $\tilde{\epsilon}$ is given by
\begin{eqnarray}
\tilde{\epsilon}_{\parallel}&=&\frac{\epsilon_1\epsilon_2}{f_1\epsilon_2+f_2\epsilon_1}\\
\tilde{\epsilon}_{\perp}&=&f_1\epsilon_1+f_2\epsilon_2, \nonumber
\label{eq:rhs}
\end{eqnarray}
 where $\epsilon_1$ and $\epsilon_2$ are the complex dielectric functions of the two constituents, respectively, and
$f_{\ell}$ are the volume fraction of the $\ell$th layer within the superlattice.

The elastic properties of superlattices are also derived from the bulk constituents
along the lines proposed by Grimsditch and Nizzoli \cite{nizzoli86}.
This approach allows to determine the total stress ($\bar{\sigma}$) and strain ($\bar{\eta}$) tensor of the superlattice
in terms of the stress and strain of the two individual layers:
\begin{equation}
\bar{\sigma}=f_1\sigma_1+f_2\sigma_2=\bar{C}\bar{\eta},
\label{eq:sigmai}
\end{equation}
where  $\bar{C}=\{\bar{c}_{ij}\}$ is the effective elastic matrix relative to the whole superlattice (see SI, Sec. S4, for the complete theoretical description). This approach has been profitably
used to study several superlattices and multilayer compounds, such as TiN/WN \cite{Buchinger:2019ds}, GaN/AlN \cite{Mohamed:2019fe}, Si/Ge \cite{Prieto:2000gz}, and Ni/Mo \cite{Martin:2005iw}.
When all the effective stiffness ($\bar{c}_{ij}$) and compliance ($\bar{s}_{ij}$) parameters are known, all the derived quantities,
such as the elastic moduli, can be obtained by substituting the effective parameters in the original formula.
The effective bulk modulus ($B$) and shear modulus ($G$) for the superlattices can be evaluated within the Voigt--Reuss--Hill approximation \cite{Kube:2016fva}, see Sec. S1 of SI for the complete formulation.
In a similar way, the Young modulus ($E_y$) and the Poisson ratio ($\nu$) as well as other mechanical properties of superlattices  can be straightforwardly derived  from the effective moduli in the Hill notation $B=B_H$ and $G=G_H$.

For example, the development of plastic deformations in crystals or the determination of micro-cracks in materials are related to the
{\em elastic anisotropy} \cite{Kube:2016fva}, which affects the mobility of dislocations.
Here, we adopted the universal anisotropic index $A_U$ proposed by Ranganathan and Ostoja-Starzewski \cite{Ranganathan:2008fw},
defined as:
\begin{equation}
A_U= 5 \frac{G_V}{G_R}+\frac{B_V}{B_R}-6  \ge 0,
\label{eq:ani}
\end{equation}
where the subscripts $V$ and $R$ identify the Voigt and Reuss expressions, respectively (Sec. S1, SI).
$A_U$  is zero for isotropic materials,
while a  deviation from zero indicates anisotropic mechanical properties.
Another relevant parameter is the Pugh modulus $G/B$, which is involved in the strain of fractures and in the
ductility of solids.
The competition between plastic flow and brittle fracture can be quantified in terms of the {\em solidity index} $S$ \cite{cottrell88}, defined as
$ S=\frac{3}{4}\frac{G}{B}$
and running from 0 and 1. The index is zero for a liquid and reaches its maximum value for diamond. $S=0.23$ is assumed as dividing point,
between ductile ($S<0.23$) and brittle ($S>0.23$) materials, which means that materials with lower solidity index show higher plasticity.
We used the Vickers model to estimate the hardness of materials \cite{fischer04}.  The Vickers scale measures the indentation
resistance which  results from the combination of resistance to plastic flow, phase transformation, and  fracturing.
The {\em Vickers hardness} $H_V$ is estimated by using the Tian empirical approach \cite{Tian:2012cz}:
\begin{equation}
H_V=0.92\Big(\frac{G}{B}\Big)^{1.137}G^{0.708}.
\label{eq:hvt}
\end{equation}
The comparison with alternative formulations of $H_V$  is reported in Table S4 of SI.
Materials can be roughly classified in term of their hardness scale \cite{Kanyanta:2016do}.
{\em Superhard materials} are defined as those with Vickers hardness greater than 40 GPa and
{\em ultrahard materials} as having a hardness exceeding 80 GPa.
Very low Vickers hardness ($H_V< 2.5$  GPa) is an unusual condition for inorganic
solid-state systems and defines the so-called {\em ultrasoft materials}.

A material is hard if it resists to indentation, i.e. to plastic deformations.
This involves the capability to contrast the nucleation and the motion of dislocations  that may generate fractures.
Thus, in the indentation hardness tests, one particular role is played by the fracture toughness ($K_{IC}$),
which measures the resistance of a material against crack propagation \cite{Feng:2011gx}.
Fracture toughness is critical for brittle materials, where the easy formation of cracks may cause to the failure of the sample.
$K_{IC}$ can be estimated combining the bulk and shear moduli as \cite{Niu:2019fd}:
\begin{equation}
K_{IC}=V_0^{1/6}G\Big(\frac{B}{G}\Big)^{1/2},
\label{eq:kic}
\end{equation}
where $V_0$ is the volume per atom.
Notably, mechanical properties such as hardness, and fracture toughness are qualitative rather than quantitative characteristics and
they depend not only on the properties of a material, but also on the measurement method and interpretation of the results  \cite{Brazhkin:2019ex}.

We initially considered superlattices with 50\%-50\% of constituent composition
(i.e. $f_1=f_2=0.5$).
In order to fulfill quality growth conditions, we imposed a maximum lattice mismatch
$\Delta a_0$ of $4\%$. This threshold allows to include systems such
as TiN/NbN ($\Delta a_0=3.9\%$) \cite{shinn1992,Kim:2011hs}, VN/TiN ($\Delta a_0=3.0\%$) \cite{Helmersson:1987fl,Musil:2000vf},
WC/TiC($\Delta a_0=1.3\%$) \cite{Zhao:2016ci}   that are stable and  successfully deposited.
Systems with larger lattice mismatch, such as TiN/CrN ($\Delta a_0=4.6\%$) \cite{YangZhao03} or VN/NbN ($\Delta a_0=6.8\%$) \cite{YangZhao03}, have been also realized, confirming the
possibility to deposit  a large range of TMX combinations. However, in the latter cases the  superlattices are plagued by a reduced thermal stability \cite{YangZhao03}. The lattice matching condition is a general pre-requisite for the growth of ordered superlattices \cite{Saha:2014gx}, while its effect on the mechanical properties is still unclear. Experimental results indicate that the difference in the elastic moduli of the single layers  plays a much more relevant role on the effective hardness of  superstructures \cite{Chu:1995kw,Chu99}.

The application of the lattice matching criterion reduces the total possible
combinations to 606 superlattices, which cover a very large range
of mechanical properties, from very brittle to very ductile or from ultrasoft to hard materials.
The complete list of elastic and mechanical properties of all 606 systems,  along with their statistical analysis are summarized in Sec. S5 of the SI.

For all multilayers, we analyzed the hyperbolicity condition for frequencies from near-IR to near-UV
($E\in[0.5-3.5]$eV). We classified a superlattice as {\em hyperbolic} if the hyperbolic condition $\epsilon_{\perp}\cdot\epsilon_{\parallel}<0$ holds at least
for a continuum energy range of 0.5 eV.  Only 113 out of the  606 ``lattice matched'' superlattices have a hyperbolic character: 38 are  type-I,
68 are type-II, and 10 are both  type-I and type-II systems depending on the frequency range (see Sec. S5, SI).
In order to find out hyperbolic metamaterials with selected mechanical hardness, we filtered the complete list of superlattices querying
for systems that are simultaneously hyperbolic and with extreme Vickers hardness:
either {ultrasoft materials} with $H_V<2.5$ GPa or {hard materials} with $H_V>18.0$ GPa.
The application of these criteria restricts the choice to 17 HMMs.

%%%%%%%%%%%%%%%%%%%%%%%%
\section{Results and discussion}
%%%%%%%%%%%%%%%%%%%%%%%%
We focus on the 17  superlattices that simultaneously fulfill the hyperbolicity and hardness conditions described above.
A few of them, such as HfN/ScN \cite{Chakraborty:2020gx,Garbrecht:2017cf}, ZrN/ScN \cite{Garbrecht:2016gqa}, TiC/TiN  \cite{Azadi:2016dt,Zhao:2010ge}, and HfC/HfN \cite{keem90}, have been  realized and experimentally characterized, although not studied  for hyperbolic optical applications.

The list of the resulting compounds and their main elastic properties are summarized in Table \ref{tab1}.
Only two out of the seventeen systems are {\em ultrasoft} materials and both include RhB as a constituent. The remaining systems are {\em hard} materials
each including at least one nitride component (mostly ScN, HfN). More generally, the most recurrent TM elements are from groups {4} and {5}, such as Zr, Nb, Hf, Ta. TMX compounds from groups {7-12} (except ZnN) do not match the selection criteria:
most of them have pure metallic character in the energy range under investigation.
All the selected superlattices do not exhibit magnetic ordering.

\begin{table}[h]
\caption{Elastic properties of hyperbolic superlattices.  Elastic moduli ($B, G, E_y$),
Young parameters (E$_y^i$) are expressed in GPa. Percent lattice mismatch ($\Delta a_0$), Poisson ratio ($\nu$), universal elastic anisotropy
(A$_U$)  are adimensional  parameters.}
\begin{center}
\begin{tabular}{c|c|c|c|c|c|c|c|c|c}
\hline
System& $\Delta a_0$& $B$ & $G$ & $\nu$ & $E_y$& $E_y^{[100]}$ & $E_y^{[001]}$ & $E_y^{[110]}$ & $A_U$    \\
\hline
\hline
RhB/RuC  &   1.3 & 310.0 &  50.1 &  0.42 & 142.6 & 400.9 & 389.5 &  83.3 & 8.20\\
RhB/ZnN  &   2.2 & 231.5 &  46.3 &  0.41 & 130.3 & 280.3 & 276.2 & 144.8 & 3.84 \\
\hline
MoB/HfN  &   1.4 & 270.9 & 160.7 &  0.25 & 402.4 & 560.8 & 560.7 & 355.6 & 0.60 \\
NbB/HfN  &   1.9 & 249.1 & 154.1 &  0.24 & 383.2 & 532.8 & 532.1 & 338.4 & 0.60  \\
NbB/ZrN  &   0.2 & 241.2 & 152.3 &  0.24 & 377.4 & 506.1 & 506.3 & 339.2 & 0.47  \\
TaB/HfN  &   1.6 & 257.6 & 160.1 &  0.24 & 397.9 & 553.9 & 554.0 & 351.3 & 0.61  \\
WC/ScN  &   2.7 & 273.2 & 167.2 &  0.25 & 416.7 & 520.0 & 473.6 & 404.7 & 0.19 \\
TaB/ZrN  &   0.2 & 249.4 & 158.3 &  0.24 & 392.1 & 527.2 & 526.4 & 352.1 & 0.48  \\
NbC/HfN  &   0.6 & 281.1 & 175.4 &  0.24 & 435.5 & 574.2 & 573.4 & 402.6 & 0.41 \\
NbC/ZrN  &   2.4 & 272.3 & 173.7 &  0.24 & 429.6 & 547.4 & 543.4 & 402.4 & 0.31 \\
HfC/HfN  &   2.7 & 248.9 & 164.7 &  0.23 & 404.7 & 502.7 & 498.6 & 383.6 & 0.25  \\
HfN/ScN  &   0.1 & 234.2 & 159.2 &  0.22 & 389.3 & 453.0 & 436.0 & 382.6 & 0.10 \\
ScN/WB  &   0.8 & 248.0 & 166.3 &  0.23 & 407.7 & 473.5 & 448.4 & 399.4 & 0.09 \\
ZrN/ScN  &   1.9 & 226.8 & 157.3 &  0.22 & 383.2 & 426.0 & 416.5 & 379.8 & 0.05 \\
TiC/TiN  &   2.0 & 259.4 & 180.5 &  0.22 & 439.6 & 489.6 & 486.5 & 426.6 & 0.06 \\
NbC/ScN  &   0.5 & 248.6 & 183.0 &  0.20 & 440.8 & 474.1 & 448.6 & 437.0 & 0.02 \\
TaC/ScN  &   0.8 & 259.0 & 193.5 &  0.20 & 464.8 & 524.3 & 478.3 & 457.1 & 0.05 \\
\hline
\hline
\end{tabular}
\end{center}
\label{tab1}
\end{table}%

\newpage
\newpage
\begin{figure}[!h!]
\begin{center}
\includegraphics[width=0.92\textwidth]{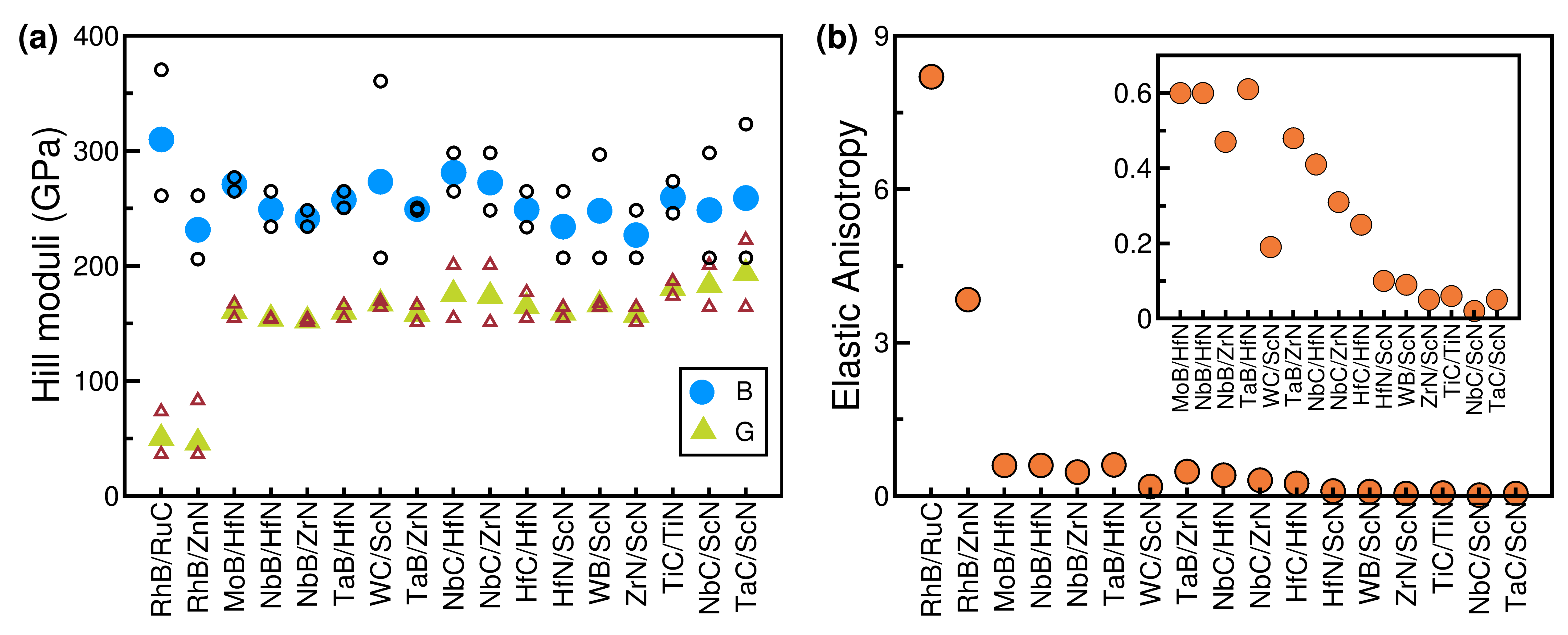}
\caption{Elastic  properties of hyperbolic superlattices: (a)  bulk $B$ and shear $G$ moduli, (b) universal
elastic anisotropy $A_U$. Full (empty) symbols correspond to superlattice (single TMX) systems. Inset in panel (b) zooms
on the lower $A_U$ values.}
\label{fig2}
\end{center}
\end{figure}
\begin{figure}[!h!]
\begin{center}
\includegraphics[width=0.92\textwidth]{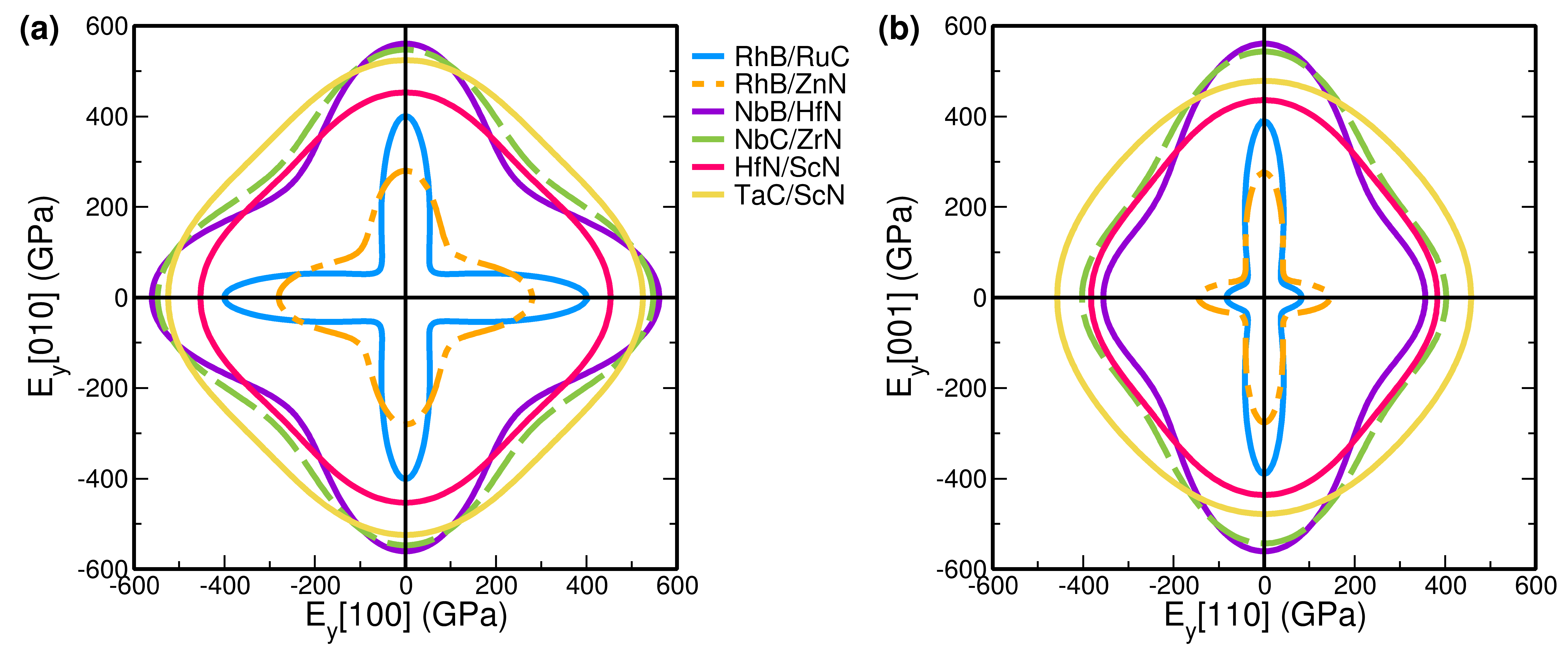}
\caption{2D plots of angular Young moduli for selected hyperbolic superlattices, projected on the  (a) (001),
and (b) ($1\bar{1}0$) planes. }
\label{fig3}
\end{center}
\end{figure}

The effective bulk and shear moduli of the hyperbolic superlattices  are shown in Figure \ref{fig2} (filled symbols), along
with the corresponding values of the single rocksalt components (empty symbols). Except for
RhB/RuC system ($B=310$ GPa), the bulk modulus of all
other compounds is in the range $B\in [200-300]$ GPa,
which reflects a high incompressibility of these materials. The shear modulus
 is instead different for ultrasoft  ($G<100$ GPa)  and hard material ($G\sim150$ GPa).
 This confirms the  predominant role of shear coefficient in the definitions of hardness (Eq. \ref{eq:hvt}). Notably,
 the value resulting for the superlattice is always included between the values of the corresponding component materials.

Figure \ref{fig2}b shows the elastic anisotropy trend: the larger is the deviation from zero, the stronger is the anisotropy.
$A_U$ index is larger than 3.8 for RhB/RuC and RhB/ZnN; while $A_U$ is less than 0.6
 for the remaining systems. This indicates that ultrasoft materials are more anisotropic than the hard ones.
The directional dependence of Young modulus is used to characterize the elastic anisotropy along crystallographic direction (Sec. S1, SI),
as  it describes the ability of a material to resist to a directional deformation
within the range applicable in linear elasticity.
Figure \ref{fig3} shows the 2D Young plots for selected hyperbolic superlattices, projected on the   (001) (panel a),
and  ($1\bar{1}0$) plane (panel b). Systems  RhB/RuC and RhB/ZnN have a lobate (flower-like) and anisotropic behavior, especially
along the [001] stacking direction.  The other systems have more symmetrical character, being almost circular (i.e. isotropic)
on  the (001) plane (Figure \ref{fig3}a), rather  all curves are prolate along the growth direction [001].
This points to the indirect interplay between  elastic anisotropy and mechanical properties \cite{Zhang:2019fm}, through the formation of dislocations in the material \cite{Sweeney:2013jn}.
The presence of dislocations increases the elastic anisotropy
and reduces the elastic moduli of the material \cite{Groh:2010fq}.

\begin{table}[h]
\caption{Mechanical and hyperbolic optical properties of hyperbolic superlattices.
Vickers hardness (H$_V$) are expressed in GPa, fracture toughness ($K_{IC}$) is expressed in MPa$\cdot$m$^{1/2}$.
Solidity index ($S$)  is adimensional.}
\begin{center}
\begin{tabular}{c|c|c|c|c}
\hline
System& $S$ & H$_V$ & K$_{IC}$&type \\
\hline
\hline
RhB/RuC  &   0.12 &   1.9 &   1.8 & II   \\
RhB/ZnN  &    0.15 &   2.2 &   1.5 & I    \\
\hline
MoB/HfN  &    0.44 &  18.5 &   3.1 & I    \\
NbB/HfN  &    0.46 &  18.9 &   3.0 & I    \\
NbB/ZrN  &    0.47 &  19.1 &   2.9 & I    \\
TaB/HfN  &    0.47 &  19.5 &   3.1 & I    \\
WC/ScN  &   0.46 &  19.7 &   3.2 & II   \\
TaB/ZrN  &    0.48 &  19.8 &   3.0 & I    \\
NbC/HfN  &   0.47 &  20.9 &   3.3 & I    \\
NbC/ZrN  &   0.48 &  21.3 &   3.3 & I    \\
HfC/HfN  &    0.50 &  21.3 &   3.1 & I    \\
HfN/ScN  &   0.51 &  21.5 &   2.9 & I+II \\
WB/ScN  &  0.50 &  21.8 &   3.0 & II   \\
ZrN/ScN  &   0.52 &  21.8 &   2.8 & I+II \\
TiC/TiN   &     0.52 &  24.1 &   3.2 & I    \\
NbC/ScN  &    0.55 &  26.0 &   3.2 & II   \\
TaC/ScN  & 0.56 &  27.5 &   3.4 & II   \\
\hline
\hline
\end{tabular}
\end{center}
\label{tab2}
\end{table}%

\begin{figure}[!h!]
\begin{center}
\includegraphics[width=1.0\textwidth]{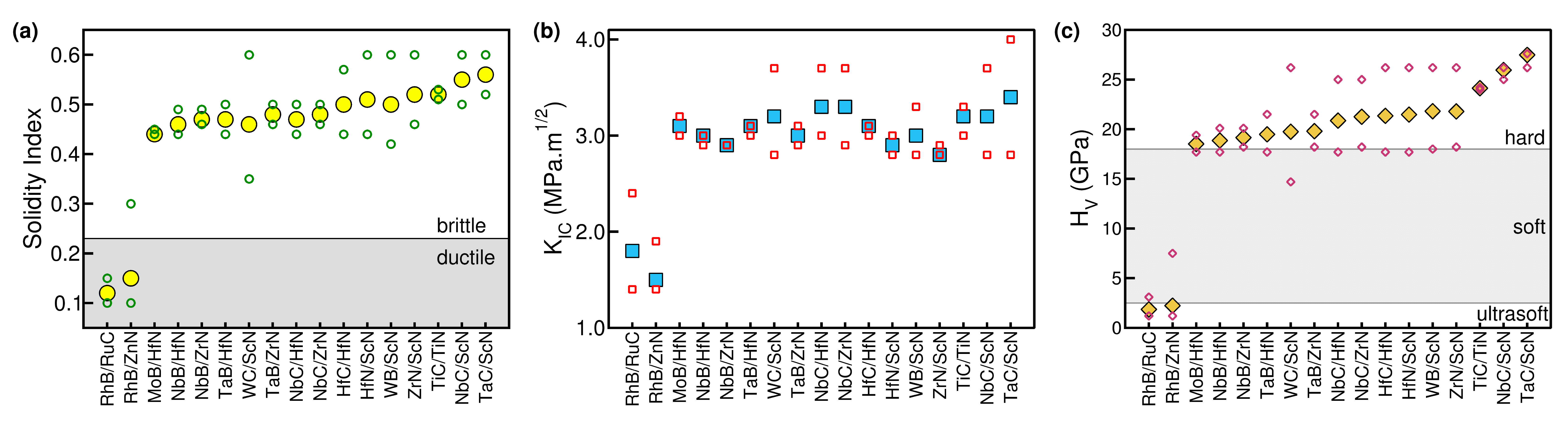}
\caption{Mechanical properties of hyperbolic superlattices: (a) solidity index $S$, (b) fracture toughness $K_{IC}$,
(c) Vickers hardness $H_V$.
Full (empty) symbols correspond to superlattice (single TMX) systems. }
\label{fig4}
\end{center}
\end{figure}

The mechanical properties of hyperbolic materials summarized in Table \ref{tab2}.
Figure \ref{fig4}a shows the solidity index $S$ (i.e. the Pugh modulus) that is used to discriminate  the brittle vs ductile behavior  of superlattices (full symbols), with respect to their
rocksalt components (empty symbols). Again, we observe a net difference between systems RhB/RuC and RhB/ZnN which are ductile ($S<0.23$) and
the remaining samples which are brittle. Except for RhB/ZnN multilayer that is composed of one brittle (ZnN) and one
ductile (RhB) element, all other systems combine TMX crystal with the same character.
The variation of the Pugh modulus is the same as that
of  the bulk and shear moduli and
is always included between the minimum and maximum values of the single components.
The $S$ coefficient is also a measure of the inhomogeneity of the electron density (the so-called {\em covalence})
and is used to discriminate the interatomic bonding character of a
crystal.  The Pugh modulus increases in parallel with the directionality of the bonds:
for metallic bonding we obtain $S<0.3$, while $S\approx0.5$ for ionic bonding and $S>0.8$ for covalent bonding \cite{Haines2011}.
In the present case, the values of systems RhB/RuC and RhB/ZnN ($S\approx0.15$)  are typical of metallic systems,
while the values of the other samples ($S\approx0.5$) indicate an ionic bonding character.

Since indentation resistance in hardness measurements is related to  both the resistance to compression (i.e. bulk modulus) and to the resistance to shape changes
(i.e. shear modulus), it is not surprising that the Pugh modulus correlates also with both fracture toughness and the Vickers hardness,
as shown in panels \ref{fig4}b and c, respectively.
Pugh modulus is negative correlation with fracture toughness: during deformation bonds break and reform resulting in displacement of atoms and slipping of atomic planes.
Materials with low fracture toughness usually exhibit high ductility and yield critical strains.
On the contrary, high Pugh modulus, which results from directional bonds, increases the shear modulus and limits the motion of dislocations that may generate fractures,
thereby increases the material hardness. Thus, RhB/RuC and RhB/ZnN are simultaneously ductile, ultrasoft, and have the lowest fracture toughness,
while all other systems are brittle, hard, with higher fracture toughness.

The calculated H$_V$ values for superlattices are in agreement  with the reduced set of available experimental data. For example, the measured microhardness for the TiC/TiN superlattice,  H$_V$=30.5-31.5 GPa \cite{Azadi:2016dt}, well matches  the theoretical value,  H$_V$=24.1 GPa, obtained with the present method.
Again, the calculated hardness of the superlattices is intermediate with respect to the single components (Figure \ref{fig4}c).
This means from one hand that the formation of multilayers does not waste the average hardness of the components, as it may happen in some alloyed mixtures \cite{Shinn:1994dr}. On  the other hand, layering cannot be assumed as a general
strategy to increase the strength of hard materials to obtain super- or ultra-hard ones.
Indeed, we do not observe  the increase of hardness ({\em hardening} effect) reported for similar systems such as TiN/NbN superlattices, for which a large range and higher hardness values have been measured: H$_V=[17.0-52.0]$ GPa
 \cite{L9_duCKh,Barnett92}. The calculated hardness for the same system is  H$_V$=16.2 GPa (see Table S3, SI) and close to the lower bound experimental limit.
The origin of such large variability and of the hardening process may depend on several
aspects \cite{Chu:1995kw,Zhang:2003bm,Subramanian:1993dx}, such as crystalline micro-granurality, grain boundaries, interface widths, cohesive forces and stress at the interfaces.
A leading role in this process is also played by dislocations and their dynamics.
Many mechanisms have been proposed to explain the dislocation effects on hardening of superlattices, including the barrier to dislocation motion provided by layers with differing shear moduli \cite{PhysRevB.2.547};
the inhibition of dislocation motion due to layer coherency strains \cite{cahn63}, or the misfit dislocation arrays at the interfaces \cite{takagi84}. This analysis goes beyond the aim of the present work and the calculated values can be safely assumed as lower  limits of superlattice hardness.

\begin{figure}[!h!]
\begin{center}
\includegraphics[width=0.85\textwidth]{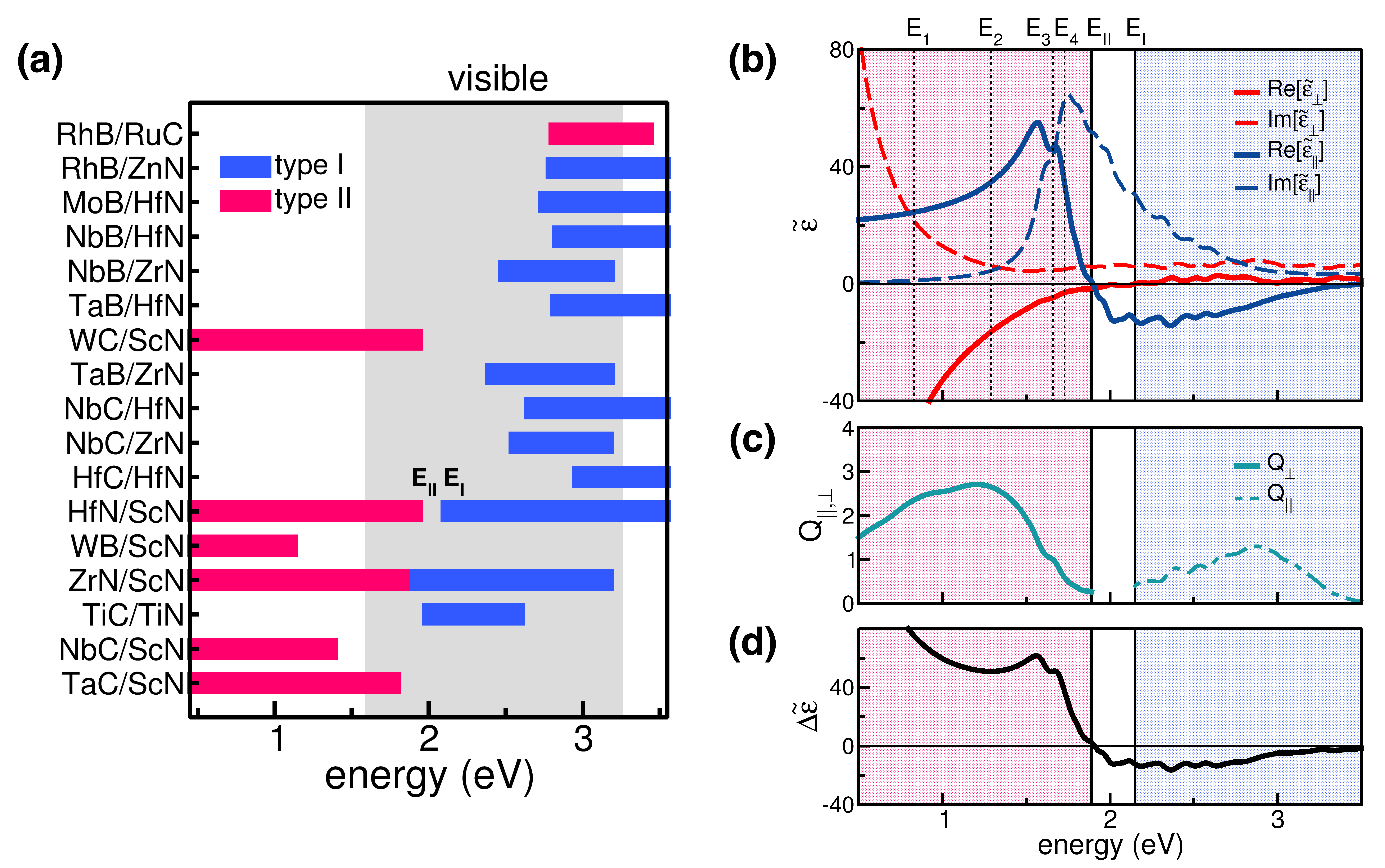}
\caption{(a) Energy distribution of hyperbolic superlattices. Gray shaded area indicates the visible range.
(b-d) Hyperbolic optical properties of HfN/ScN superlattice:
(b) real and imaginary part of the effective
dielectric functions $\tilde{\epsilon}_{\parallel}$ and $\tilde{\epsilon}_{\perp}$;
 (c) parallel  and perpendicular  quality factors $Q_j$; and (d) strength of dielectric anisotropy $\Delta\tilde{\epsilon}$.
  Shaded areas in panel (b-d) indicate the energy regions with type-II ($E<E_{II}$, red area) and type-I ($E>E_I$, blue area) hyperbolic behavior.
Vertical dotted lines in panel (b) mark selected energies $E_i$, discussed  in the angular analysis of the dielectric function
$\epsilon_{\varphi}$ (see Figure \ref{fig6}).}
\label{fig5}
\end{center}
\end{figure}

The hyperbolic character of the superlattices and their specific operational window is shown in Figure \ref{fig5}a, where we focus in the energy
range $E\in[0.5-3.5]$ eV, which includes  near-IR,  visible, and near-UV range that are the most interesting for optical applications.
RhB/RuC and RhB/ZnN superlattices are hyperbolic in the visible-UV part of the spectrum, all other systems  exhibit a type-I character at higher
energies or a type-II character at lower energies.

We investigate in more details the optical properties of HfN/ScN,
that we assume as the reference for this class of TMX multilayers. HfN/ScN has been experimentally grown  \cite{Chakraborty:2020gx,Garbrecht:2017cf} and behaves as both type-I and type-II material, depending
on the energy of the incoming radiation. Figure \ref{fig5}b shows the real and  imaginary part of both parallel $\tilde{\epsilon}_{\parallel}$ and perpendicular $\tilde{\epsilon}_{\perp}$ components of the effective
dielectric function (Eq. \ref{eq:rhs}); see Fig. S2 in SI for the details of the constitutive dielectric functions. We define $E_I$ and $E_{II}$, the energies at which the system changes its optical behavior from regular (e.g. metal) to hyperbolic type-I ($E_I$) or type-II ($E_{II}$), respectively.
The parallel component (blue lines) has the  behavior typical of semiconductors:
the real part (straight line) is always positive for $E<E_{II}=1.89$ eV and reaches the dielectric constant value ($\epsilon_0$ = 21.8) in the limit
for $E\rightarrow0$. The imaginary part (dashed line) has a peak at $E=1.76$ eV, which mainly derives from interband optical transitions in the ScN part.
The perpendicular component (red lines) has, instead, a metallic character: for $E<E_I=2.14$ eV the real part is negative and diverges for $E\rightarrow0$, with a
typical Drude-like behavior that mostly comes from the intraband transitions of HfN component.
 Thus, for $E<E_{II}$  $Re[\tilde{\epsilon}_{\parallel}]$ is positive and $Re[\tilde{\epsilon}_{\perp}]$ is negative, which corresponds to a
type-II system. The opposite happens for $E>E_{I}$, where $Re[\tilde{\epsilon}_{\parallel}]<0$ and $Re[\tilde{\epsilon}_{\perp}]>0$, and
the system has a type-I character. In the intermediate range between $E_{II}$ and $E_{I}$, both components are negative and the system is metallic.

In order to quantify the hyperbolic character of the multilayers, we define two figures of merit \cite{Hoffman:2007jt,Drachev:2013gn},
namely the quality factor $Q_{j}=-\frac{Re[\tilde{\epsilon}_j]}{Im[\tilde{\epsilon}_j]}$, with $j=\parallel, \perp$; and the
strength of dielectric anisotropy  $\Delta\tilde{\epsilon}=Re [\tilde{\epsilon}_{\parallel}-\tilde{\epsilon}_{\perp}]$.
The quality factor $Q$ accounts for the energy losses only in the direction in which the material has a metallic character.
Systems with the quality factor  $Q\approx 3$ or higher
are considered good HMMs \cite{Korzeb:2015fv}.
The results (Figure \ref{fig5}c) indicate that the quality factor reaches
the maximum at $E\sim$1.20 eV where $Q_{\perp}$=2.8, this qualifies HfN/ScN as a hyperbolic material
with reduced energy loss in the near-IR range. At higher energies ($E>2.0$ eV) the Q factor is almost
halved, indicating a higher metal-like dissipation. The role of losses has to be taken with care and depends on specific
optical needs: while for systems  based on energy transport (such as  waveguides), life time of excitations
has to be maximized and energy losses have to be reduced as much as possible,
for other applications (such as thermal absorbers) which are based on plasmon decay, effects associated with inelastic dissipation are highly desired.
The $\Delta\tilde{\epsilon}$ factors quantifies the metal-dielectric anisotropy along the direction parallel and perpendicular to the
optical axis. Systems with $\Delta\tilde{\epsilon} >20$ are considered good
hyperbolic materials \cite{Korzeb:2015fv,Ishii:2013er}. In the case of HfN/ScN, this condition is fulfilled almost in
the entire range $E<E_{II}$, where the system has a type-II behavior.

The energy range, the hyperbolic type, and the gain factors strictly depend on the composition (i.e. the
crossover energies, and optical transitions in each TMX layer) and on the thickness ratio (e.g. $f_i$) of
the multilayer, as shown in Figure S8 of SI, where the gain factors of selected
superlattices are shown. Ultrasoft systems RhB/RuC and RhB/ZnN have low quality  factors ($Q<1.5$) and
low dielectric anisotropy ($\Delta\tilde{\epsilon}<20$) for the largest part of the energy spectrum under consideration.
We can conclude that RhB/RuC and RhB/ZnN conjugate rare qualities such as ultrasoft hardness and hyperbolic light dispersion, but are
plagued by high energy dissipation because of interband transitions from non-metal $sp$  to  transition-metal $d$ bands.
Hard superlattices HfN/ScN, ZrN/ScN and TaC/ScN have instead the highest gain factors that correspond to smaller
energy loss. Depending on the specific need of an application - such as the energy
working range or the dissipation tolerance -  it is possible to
select the optimal TMX pair and to fine tune the chemical and structural details
of the superlattice.
The strength of dielectric anisotropy also affects the dispersion relation of the  radiation that
may propagate along the medium, as $\epsilon_{\parallel}$ and $\epsilon_{\perp}$ define the
geometric characteristics of the hyperbolic isosurface equation:
\begin{equation}
\frac{k_1^2+k_2^2}{\epsilon_{\parallel}}+\frac{k_3^2}{\epsilon_{\perp}}=\frac{k_{\perp}^2}{\epsilon_{\parallel}}+ \frac{k_{\parallel}^2}{\epsilon_{\perp}}=\frac{\omega^2}{c^2}.
\label{kdi}
\end{equation}

\begin{figure}[!h!]
\begin{center}
\includegraphics[width=0.9\textwidth]{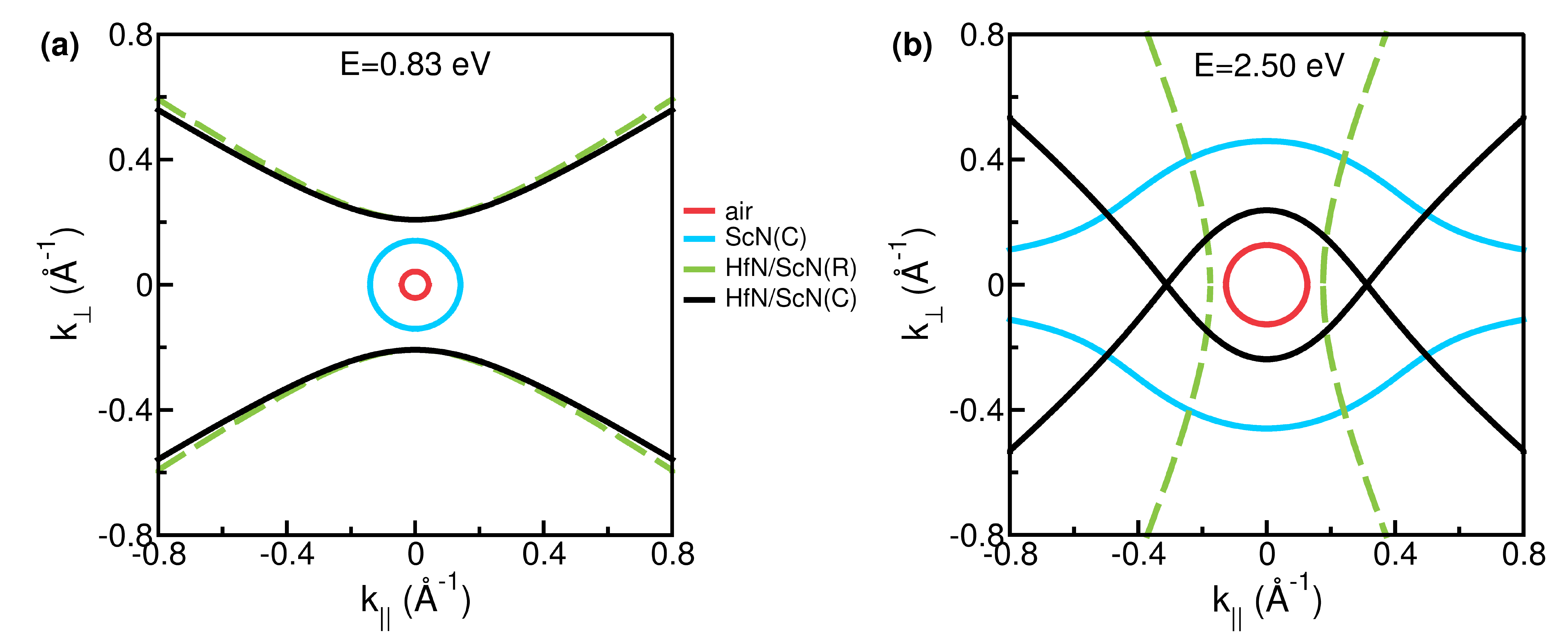}
\caption{2D k-dispersion plot for dielectric (air and ScN) and hyperbolic (HfN/ScN) materials at (a) $E=0.83$ eV, and (b) $E=2.50$ eV.  Labels (R) and (C) indicate if only real part or the complete complex dielectric function is considered in Eq.\ref{kdi}.}
\label{fig6}
\end{center}
\end{figure}
The k-dispersions for HfN/ScN case are shown in Figure \ref{fig6}, along with the plots for representative dielectric systems (namely air and ScN), included for comparison. We considered  two radiation energies
$E=0.83$ eV (panel a) and $E=2.50$ eV (panel b) that are representative for type-II and type-I character of HfN/ScN superlattice.
The {\bf k} wavevector follows a truly hyperbolic surface (i.e. open form) only for the ideal case with zero absorption processes,
i.e. when only the real part of the dielectric function is considered in Eq. \ref{kdi}.
In the case of absorbing materials, the isosurface equation assumes a form whose final dispersion
crucially depends on the imaginary part of the dielectric function.
In the case of air, the dielectric function reduces in all direction to the dielectric constant $\epsilon_{r}=1.0$, and the  k-dispersion relation is a circumference for both energies.
The case of ScN at $E=0.83$ eV is similar: the imaginary part of the dielectric function is almost zero.
At higher energy ($E=2.50$ eV) ScN is optically active and the imaginary part plays a crucial role ($\epsilon_{r}=11.39$, $\epsilon_{i}=9.57$). The corresponding k-isosurface has a prolate open shape (panel b).
For HfN/ScN superlattice at $E=0.83$ eV (type-II material) the inclusion of the imaginary part has minor effect on the
k-dispersion. In this case the hyperboloid isosurfaces reduce to a hyperbola (Figure \ref{fig6}a).
At $E=2.50$ eV (type-I material) the hyperbola is open if only the real part of the dielectric function is considered. The inclusion of the imaginary part (i.e. dissipation) gives rise to a close loop (panel b).

Since electromagnetic radiations propagate through dielectric systems ($\epsilon_r>0$), while
cannot travel across metals ($\epsilon_r<0$), the hyperbolicity condition $\epsilon_{\perp}\cdot\epsilon_{\parallel}<0$
opens to the question on wether and under which conditions an electromangnetic wave may propagate through
layered metamaterials. In single planar metal-dielectric interfaces, under proper conditions, it is possible to excite
surface plasmon-polariton waves that travel along the interface.
The case of hyperbolic superlattices is more complex because
they act as a metal in one direction and as a dielectric in the other.

\begin{figure}[!h!]
\begin{center}
\includegraphics[width=0.7\textwidth]{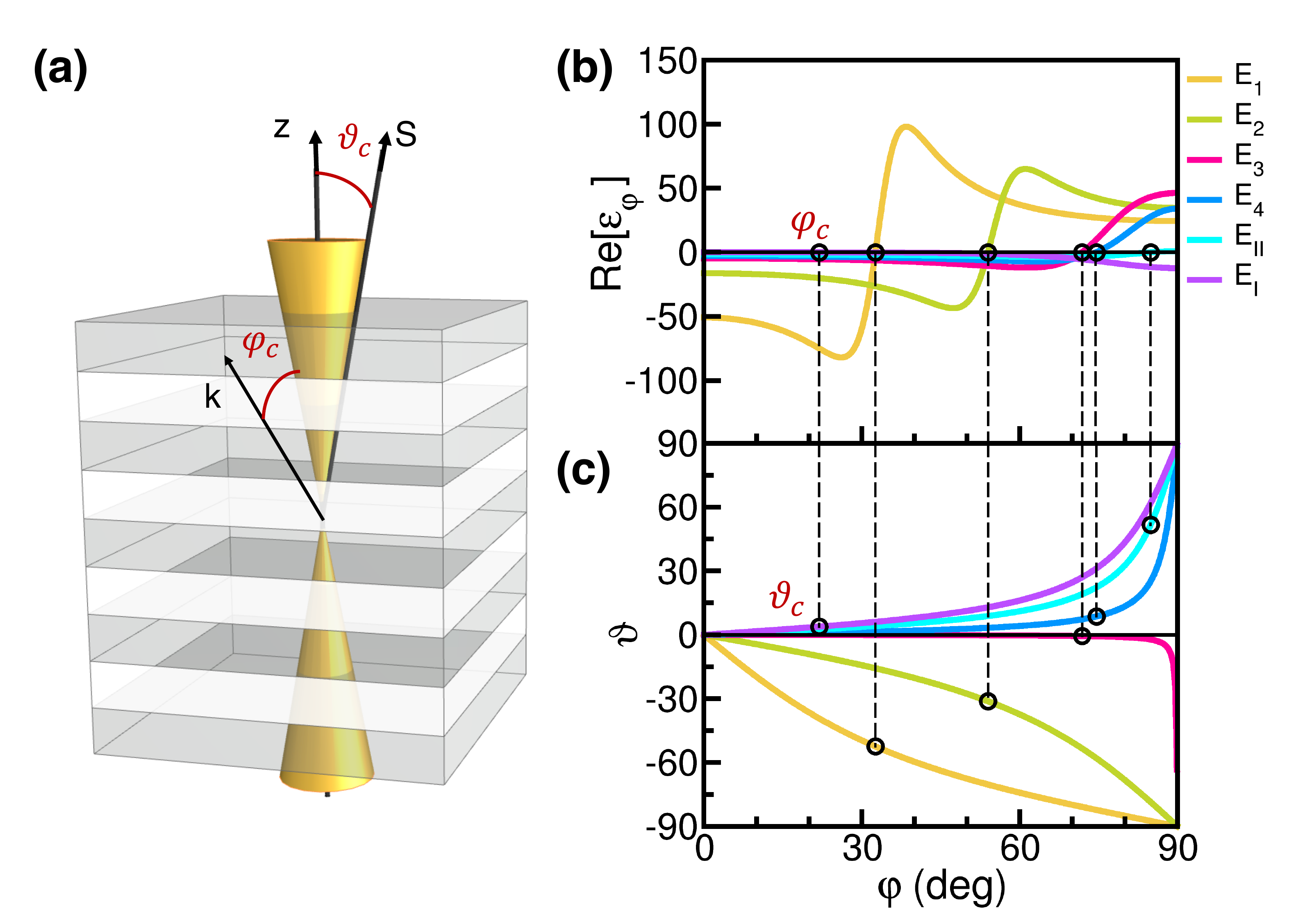}
\caption{ (a) Vector diagram for transverse magnetic propagating waves (TM) and hyperbolic dispersion isosurface corresponding to a type-II HMM;
(b) Real part of the effective function $\epsilon_{\varphi}$ and (d) $\vartheta(\varphi)$ angle between the extraordinary
wave and the optical axis, at different energies ($E_i$) of the incoming electric field, for the HfN/ScN superlattice.
Energies ($E_i$) are defined in Figure \ref{fig5}b. Vertical dashed lines identify the critical angles
$\varphi_c$ and $\vartheta_c$, respectively.}
\label{fig7}
\end{center}
\end{figure}

Hyperbolic superlattices can be considered as plasmon-polaritonic crystals where the coupled states of light and electron density (i.e. plasmons) give rise to traveling extraordinary waves \cite{Lindell:2001cu}, known as {\em volume plasmon polaritons} (VPPs)  \cite{Zhukovsky:13,Ishii:2013er}.
In anisotropic materials, the electric-field vector {\bf E} and  the electric-displacement vector {\bf D} are not usually parallel.
As a consequence, the Poynting vector {\bf S} and the direction of the wavefront {\bf k} are no more parallel (Figure \ref{fig7}a).

The coupling between the Poynting vector and the wavevector can be investigated
through the angular description of the dielectric function \cite{Ishii:2013er}:
\begin{equation}
\frac{1}{\epsilon_{\varphi}}=\frac{sin^2\varphi}{\tilde{\epsilon}_{\parallel}}+\frac{cos^2\varphi}{\tilde{\epsilon}_{\perp}},
\label{eq:phi}
\end{equation}
where $\varphi$ is the angle between the wavevector of the radiation  {\bf k} and the optical axis {\em z} (Figure \ref{fig7}a).
Since $\tilde{\epsilon}_{\parallel}$ and $\tilde{\epsilon}_{\perp}$ are functions of the energy of the incoming radiation,
$\epsilon_{\varphi}$ is also a function of the energy.
Figure \ref{fig7}b shows the real part of $\epsilon_{\varphi}$ of the HfN/ScN system, for a set of selected
energies, $E_i$,  marked as vertical lines in Figure \ref{fig5}b.
$E_I$ and $E_{II}$ are the energies that delimit the type-I and type-II ranges, respectively and are both in the visible range.
$E_1=0.83$ eV ($\lambda=1.5\ \mu$m) is characteristic energy (wavelength) for telecommunications, $E_2=1.29$ eV is a representative
energy in the near IR, $E_3=1.66$ eV and $E_4=1.73$ eV are selected energies in the visible range.
For all energies, except $E_I$, $Re[\epsilon_{\varphi}]$ has the same general trend, being negative at low angles and positive for $\varphi\rightarrow\pi/2$,
the opposite holds for $E_I$. For $\varphi=0$ ($\varphi=\pi/2$) $\epsilon_{\varphi}=\tilde{\epsilon}_{\perp}$ ($\epsilon_{\varphi}=\tilde{\epsilon}_{\parallel}$) that is
negative (positive) for a type-II (type-I) HMM.
As the energy increases from $E_1$ to $E_I$, $Re[\epsilon_{\varphi}]$ we observe a reduction of the intensity especially of the negative part, along with a shift of the maxima/minima towards higher angles.

The condition $Re[\epsilon_{\varphi}]=0$ determines the angular boundary between the metallic and dielectric response of the metamaterial,
which is the analogous of plasmon excitation in planar interfaces.
The critical angle $\varphi_c$ is the direction of radiation at which there can be the  excitation of a travelling plasmon-polariton wave in the system.
If we consider the interface with another dielectric medium, the wavevector at the interface follows the Snell's law.
Thus, the angle $\theta$ between the extraordinary wave (i.e. the Poynting vector) and the optical axis is:
\begin{equation}
\vartheta(\varphi)=arctan \bigg( Re[\frac{\tilde{\epsilon}_{\parallel}} {\tilde{\epsilon}_{\perp}}] tan(\varphi) \bigg).
\label{eqtheta}
\end{equation}
At the critical angle $\varphi_c$,  the quantity $\theta_c$  represents the angle of propagation of the VPP along the metamaterial  \cite{Zhukovsky:2013if}.
This means that VPP  propagates  throughout the {\em  volume} of the HMM along a {\em cone}
with axis coincident to the optical axis and aperture $\theta_c$ (Figure \ref{fig7}a).
The trend for $\varphi_c$ follows the one observed for $Re[\epsilon_{\varphi}]$, as shown in Figure \ref{fig7}b.
The energy dependent  $\vartheta(\varphi)$ and the critical angles $\vartheta_c$ are shown in panel (c).
We observe a change in the concavity of the $\vartheta(\varphi)$ as a function of the incoming radiation energy (e.g. $E_1$ and $E_2$).
For  $E\le E_3$, $\vartheta_c$ is negative  implying an anomalous refraction  for the extraordinary wave.
The refraction is anomalous in the sense that the transmitted wave is a {\em backward wave} with the Poynting vector and
the wavevector that are in opposite lateral directions.
This is not a case of double negative refraction
as originally formalized for left-handed materials by  Veselago \cite{Veselago:67}, i.e. negative dielectric permittivity $\epsilon<0$ and negative magnetic permeability
$\mu<0$.
Rather, it has been shown that it possible to obtain {\em backward waves} with uniaxial anisotropic media, under certain conditions and
 when only one parameter has a negative value \cite{Lindell:2001cu}.
In the present case, $\mu_{\parallel},\mu_{\perp},\epsilon_{\parallel}>0$ and $\epsilon_{\perp}<0$. This allows for backward waves  only for transverse magnetic mode radiation \cite{Lindell:2001cu}.
For $E>E_3$, $\vartheta$ is positive, even though (as for $E_4$ and $E_{II}$) the system as the same type-II character.
This is due to the $\tilde{\epsilon}_{\parallel}/\tilde{\epsilon}_{\perp}$ fraction in Eq. \ref{eqtheta}.
For $E>E_I$ both $\varphi_c$ and $\vartheta_c$ are quite zero and positive, and  $\epsilon_{\varphi}\simeq\tilde{\epsilon}_{\perp}$
is small and positive (Figure \ref{fig5}b). The excited transverse magnetic mode waves can thus travel almost parallel to the optical axis as through a
low dielectric medium. This analysis opens up the possibility to use hard materials also for negative refraction applications and cloaking in the visible range.

 \begin{figure}[!h!]
\begin{center}
\includegraphics[width=0.92\textwidth]{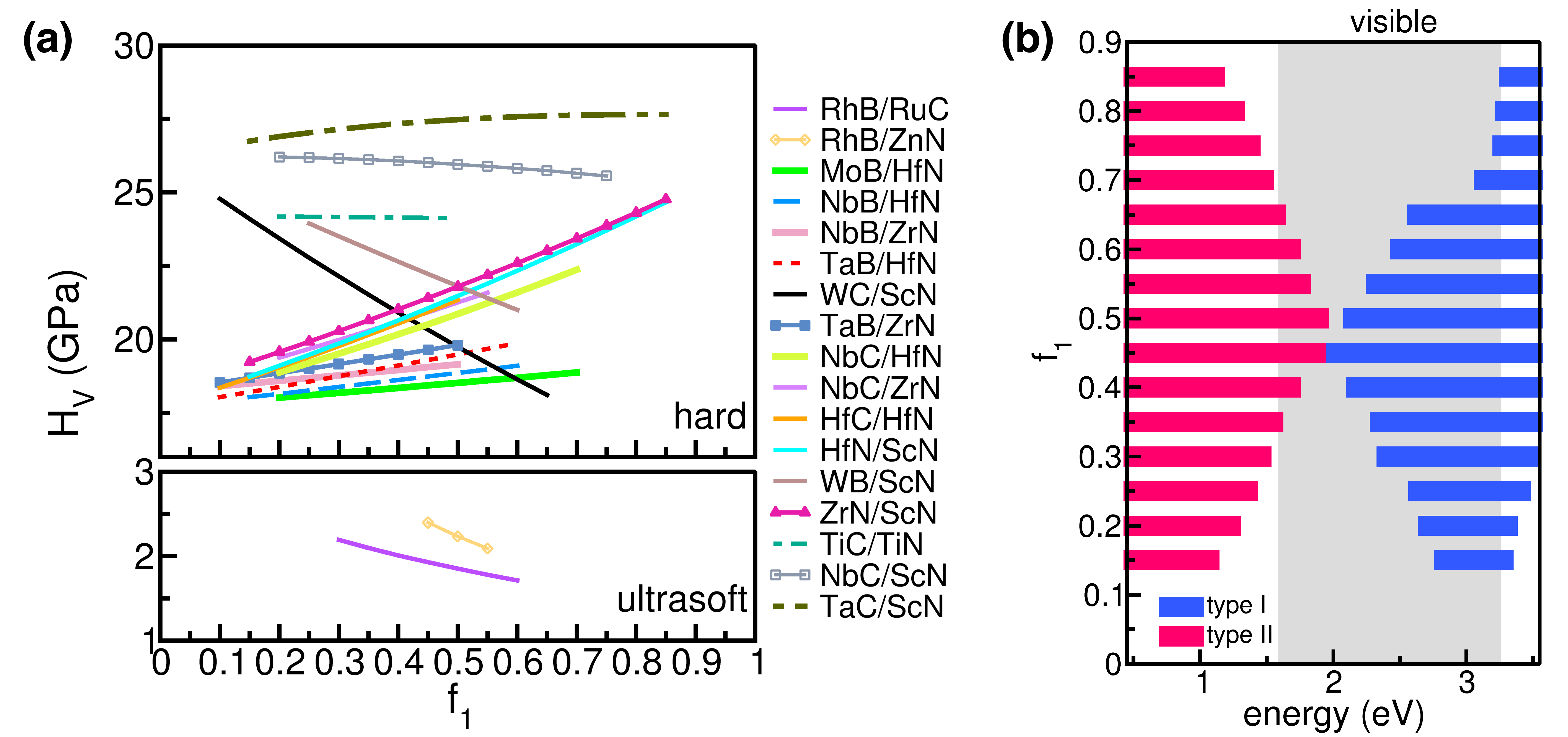}
\caption{(a) Hardness of hyperbolic superlattices as a function of the filling factor $f_1$.
(b) Energy distribution of hyperbolic response for HfN/ScN, as a function of the filling factor $f_1$.}
\label{fig8}
\end{center}
\end{figure}

Finally, we investigated the effect of the filling factor $f_{\ell}$ in the results presented above.
For the seventeen selected superlattices we calculated the mechanical properties and the
hyperbolic behavior for $f_1$ ranging from 0 to 1 ($f_2=1-f_1$). In the case, e.g. of RhB/RuC pair,
$f_1=0$ corresponds to the RuC single crystal and $f_2=1$ corresponds to the RhB one, the same for all the other systems.
The results for the calculated Vickers hardness are shown in Figure \ref{fig8}a.
The volume ratio $f_{\ell}$ enters in the definition of both the effective stiffness coefficients of the superlattices and the
effective parallel and perpendicular dielectric functions. Thus, the selective criteria for the hardness or hyperbolic behavior
do not necessarily hold for all the $f_{\ell}$ values. For example, in the case of RhB/ZnN system, all the criteria are fulfilled only in the range
$f_1\in[0.45-0.55]$. In general, the hardness of superlattices increases  as the volume of the harder layer is increased.
For the majority of the systems, this requires a reduction of the nitride components.
In the particular case of HfN/ScN and ZrN/ScN, that are composed two nitride species,
the hardness increases as the  less hard layers (HfN and ZrN) become thinner.
WC/ScN and WB/ScN superlattices have the opposite trend.
Increasing the hardness has a price and affects the optical response of the HMMs.
This is evident from Figure \ref{fig8}b, which displays the energy  distribution of the hyperbolic response for HfN/ScN, as a function of the filling factor $f_1$. The largest energy range for hyperbolic optical responses is close to $f_1=0.5$, while both type-I and type-II ranges reduce
as the layer composition moves from 50-50\%. In other cases, as for RhB/RuC, NbB/HfN, NbB/ZrN, TaB/HfN and TaB/ZrN, the hyperbolic character
changes with the volume composition: RhB/RuC is type-I for $f_1<0.45$ and type-II for $f_1\ge0.45$; NbB/HfN is type-I in the vis-UV region for the entire
range $f_1\in[0.15-0.60]$ and it is also type-II in the near-IR part of the spectrum for $f_1\in[0.25-0.35]$. Finally, other rocksalt pairs that
do not match the selection criteria for $f_{\ell}=0.5$,  fulfill the query for different volume ratio, for example at $f_1=0.75$  AgC/ScN and CuC/RuC
superlattices are ultrasoft materials ($H_V$=2.0, and 2.3,  GPa) with type I+II  and type-II  hyperbolic character, respectively.
We conclude that, within the limits of the effective medium approach, the volume ratio is a critical degree of freedom that can be  tailored
to reach the specific mechanical and optical properties needed for a specific application.

%%%%%%%%%%%%%%%%%%%%%
\section{Conclusion}
%%%%%%%%%%%%%%%%%%%%%

The realization of HMM relies on the fabrication of multilayer structures where the choice of the component materials provides the strategy to fine engineering the optical and mechanical properties.
We used high-throughput techniques based on density functional and effective medium theory
to design HMMs with selected mechanical properties, by combining the properties of simple TMX rocksalt crystals.
The simultaneous requirement of structural lattice match, extraordinary mechanical hardness (both solid-state ultrasoft and hard
materials) and hyperbolic optical response restrict the wide range of possible combinations to 17
superlattices, out of 1891 combinations, most of which include nitride components. A few compounds include quite exotic crystal structures such as RuC and ZnN, or monoborides whose
stability has to be confirmed by experiments. The majority  of the systems are composed of transition metal nitrides and carbides (e.g. HfN, TiN, ZrN, NbC, TiC, TaC) that are
easily grown and whose stability is largely demonstrated. A few of the proposed superlattices, such as HfN/ScN, ZrN/ScN,
seem to be particularly promising: they have been experimentally realized, and are expected to be simultaneously mechanically hard and optically hyperbolic over a large range of frequencies. This calls for further experimental  validations and characterizations.
The choice of the material layers and their relative thickness can be
fine tuned to fulfill {\em ad hoc} mechanical and optical requirements to optimize the performance of the final
optoelectronic application. This opens up ways for the design and growth of hyperbolic materials with superior mechanical
properties that can be used in extreme conditions as for aerospace or security applications.

% Experimental section

\section{Methodology and Computational Details}
Electronic structure calculations were carried out by using a first-principles total-energy-and-forces
approach based on density functional theory (DFT), as implemented in the QUANTUM ESPRESSO (QE) suite of codes \cite{Giannozzi:2009p1507}.
The Perdew, Burke, and Ernzerhof (PBE) \cite{pbe} generalized gradient approximation was adopted for parametrization of the
exchange-correlation functional. Ionic potentials were described by normconserving Trouillier-Martin pseudopotentials.
Single particle Khon-Sham orbitals are expanded in plane waves up to a kinetic energy  cutoff of 150.0 Ry.
A uniform (24 $\times$ 24 $\times$ 24) k-point grid is used for summations over the Brillouin zone.
Each system has a rocksalt structure (NaCl type, space group $Fm\bar{3}m$, number 225), with two inequivalent atoms in the primitive cell. The equilibrium lattice constance (a$_0$) of each system is obtained through a variable cell optimization  \cite{vc-relax}.
See Sec. S1 of SI for further computational details. 

In the case of semiconducting compounds (ScN, Yn, LaN), the electronic properties
have been calculated ab initio by using a recent pseudohybrid Hubbard implementation of DFT+U, namely ACBN0 \cite{acbn0}, that profitably corrects  the energy
bandgap \cite{acbn0,Gopal:2015bf} as well as the dielectric and vibrational properties of semiconductors \cite{Calzolari:2013kv}.
The optimized values for the studied compounds  are U$_N$(ScN)=3.1 eV, U$_{Sc}$(ScN)=0.1 eV, U$_N$(YN)=3.10 eV, U$_{Y}$(YN)=0.1 eV,
U$_{N}$(LaN)=3.3 eV, U$_{La}$(ScN)=0.0 eV.
The spin degrees of freedom for magnetic compounds (MnB, LaB, CrC, MnC, FeC, CrN, MnN, FeN)
have been described within the local spin-density approximation (LSDA). Preliminary tests confirmed that spin-orbit coupling (SOC) contributions
give negligible corrections to the ground state of single crystals in the {\em fcc} structure.

The optical properties of rocksalt crystals are
determined using the epsilon.x code, contained in the QE package, which implements  a band-to-band independent-particle (IP) formulation
of the frequency-dependent Drude-Lorentz model for the dielectric function $\hat{\epsilon}(\omega)$, where both intraband
(Drude-like) and interband (Lorentz-like) contributions are explicitly considered \cite{Calzolari:2014gj}:
\begin{equation}
\hat{\epsilon}(\omega)= 1 - \sum_{{\bf k},n}f_{{\bf
k}}^{n,n}\frac{\omega_p^2}{\omega^2+i\eta\omega} + \sum_{{\bf
k},n\neq n'}f_{{\bf k}}^{n,n'}\frac{\omega_p^2}{\omega^2_{{\bf
k},n,n'}-\omega^2-i\Gamma\omega}, \label{eps}
\end{equation}
where
$\omega_{p}$ is the bulk plasma frequency; $\hbar\omega_{{\bf k},n,n'}= E_{{\bf
k},n}-E_{{\bf k},n'}$ is the vertical band-to-band transition
energy between occupied  and empty  Bloch states calculated at the DFT level;
 $\{{\bf k},n\}$  and $\{{\bf k},n'\}$ are the k-point and band  quantum numbers, respectively.
 $\eta,\Gamma \rightarrow 0^+$ are the Drude-like and Lorentz-like
relaxation terms, while $f_{{\bf k}}^{n,n}$ and $f_{{\bf
k}}^{n,n'}$ are the corresponding oscillator strengths within the dipole approximation (see
S.I. for further details). 

Despite the simplicity, the capability of the present approach in simulating the optical properties of plasmonic materials has been extensively proved, e.g., in Refs \cite{Calzolari:2014gj,Catellani:2020fj,Miller2018}.
In the particular case of TiN,  the accuracy of this approach has been tested in Refs. \cite{Catellani:2017gza} and \cite{Shah2018},
where the dielectric function of TiN bulk calculated with the Drude-Lorentz approach is compared to one based on Time-Dependent Density Functional Perturbation Theory and to the experimental data  \cite{Pfluger:1984jn,Herzing:2016io}.
A similar Drude-Lorentz approach has been profitably used by other authors for simulating the dielectric function of metallic rocksalts
(e.g. ZrN, HfN, TaC, WC) \cite{Kumar:2015hna}, in very good agreement with the present results.
In the case of ScN (see SI, Sec. S2), the comparison with experimental results \cite{PhysRevB.63.125119} confirms the accuracy of the present approach also in the case of small gap semiconductors.

Second order elastic stiffness (c$_{ij}$) and compliance (s$_{ij}$) coefficients have been calculated by means of
a polynomial fitting of the strain-stress relation of deformed crystals, as implemented in the ElaStic simulation package  \cite{Golesorkhtabar:2013jy}.
For each rocksalt crystal, we considered 11 distorted structures with a Lagrangian strain
value between  $-\eta_{max}$ and  $+\eta_{max}$, with $\eta_{max}=1\times10^{-3}$. The symmetry-dependent
deformation types follows the universal linear-independent coupling strains (ULICS) classification,
proposed in Ref. \cite{Yu:2010dz}. The energies and stresses of the distorted structures are calculated ab initio
by using the QE code (see
S.I. for further details). 

High throughput DFT, optical, and ElaStic calculations have been run by using the automatic workflows
implemented in the AFLOW$\pi$  infrastructure \cite{aflowpi}.

\medskip
\textbf{Supporting Information} \par %Please delete the Suppporting Information statement if it is not applicable. Please supply Supporting Information in another file. Supporting information should not be provided in .tex format
Supporting Information is available from the Wiley Online Library or from the author.
The Supporting Information file includes: the description of the DFT implementation of the optical and elastic properties
of crystals (Sec. S1);  the summary of the main structural, electronic and optical properties of single TMX systems (Sec. S2); the complete list of the elastic and mechanical properties of single rocksalt crystals (Sec. S3); the detailed description of the mean theory approach for the evaluation of the elastic properties in superlattices (Sec. S4); the complete list of the elastic and mechanical properties of all TMX-based superlattices (Sec. S5); and the description of the
gain factors (Sec. S6) discussed in the text.

% Acknowledgements

\textbf{Acknowledgements} \par %delete if not applicable))
We acknowledge the support of the High Performance Computing Center at the University of North Texas and the Texas Advanced Computing Center at the University of Texas, Austin.
The authors thank Rita Stacchezzini for graphical help as well as Andrew Supka and
Sharad Mahatara for technical support and discussions.

% References

%
\bibliography{biblio}

\end{document}